\documentclass[prl,floatfix,showpacs,amsmath,twocolumn]{revtex4}
\usepackage{amssymb}
\usepackage{graphicx}
\usepackage{amsmath}
\usepackage{amsthm}

\newcommand{\bea}{\begin{eqnarray}}
\newcommand{\eea}{\end{eqnarray}}
\def\bi{\begin{itemize}}
\def\ei{\end{itemize}}
\def\bc{\begin{center}}
\def\ec{\end{center}}

\def\C{\hbox{$\mit I$\kern-.7em$\mit C$}}
\def\R{\hbox{$\mit I$\kern-.6em$\mit R$}}

\def\ket#1{|#1\rangle}

\def\ket#1{\left| #1\right>}

\def\et{\eta}
\def\etd{\eta^{\dagger}}

\def\upa{\uparrow}
\def\dna{\downarrow}
\def\cdag#1{c_{#1}^{\dagger}}

\def\cond{|\eta_N\rangle}

\def\dnoc{\langle\eta_N|}

\begin{document}

\title{An $\eta$-condensate of fermionic atom pairs via adiabatic state preparation}

\author{A. Kantian}
\author{A. J. Daley}
\author{P. Zoller}
\affiliation{Institute for Theoretical Physics, University of
Innsbruck, A-6020 Innsbruck, Austria\\ and Institute for Quantum
Optics and Quantum Information of the Austrian Academy of Sciences,
A-6020 Innsbruck, Austria}

\begin{abstract}
We discuss how an $\eta$-condensate, corresponding to an exact
excited eigenstate of the Fermi-Hubbard model, can be produced with
cold atoms in an optical lattice. Using time-dependent density
matrix renormalisation group methods, we analyse a state preparation
scheme beginning from a band insulator state in an optical
superlattice. This state can act as an important test case, both for
adiabatic preparation methods and the implementation of the
many-body Hamiltonian, and measurements on the final state can be
used to help detect associated errors.
\end{abstract}

\maketitle

Experiments with cold atoms in optical lattices not only make
possible the realisation of many-body lattice Hamiltonians and their
corresponding ground states~\cite{Reviews,ExpOverview}, but also
exhibit long coherence times. This opens the way to produce excited
many-body states and consider the related quantum dynamics, as
demonstrated by recent investigations of repulsively bound atom
pairs~\cite{RepPairs1,RepPairs2}. A key question in this context is
how to prepare specific excited states, especially those
corresponding to interesting quantum phases. Here we show that exact
excited eigenstates of the Fermi Hubbard model, the
$\eta$-condenstates first discussed by Yang~\cite{Yang} can be
realised in experiments by combining an adiabatic ramp beginning
from an insulating state in an optical superlattice with a sudden
switch in the interaction strength (see Fig.~1a). These states
exhibit long range order in all dimensions and have been discussed
in the context of high temperature
superconductivity~\cite{DemlerZhangReview}. Moreover, as exact
excited eigenstates they provide (i) an ideal test case for the use
of adiabatic ramping processes in state
preparation~\cite{Loading1,Loading2,RoschVojta}, which has important
possible applications in the production of low-entropy ground
states, and (ii) the possibility to validate the implementation of
the many-body Hamiltonian, by testing the properties of the final
state.

Below we show that the state preparation
process proceeds with high fidelities for realistic experimental
size scales and parameters, even in the presence of imperfections and noise.
We focus on the 1D case, where time-dependent density matrix renormalisation group (TDMRG)
methods~\cite{DMRGreview} allow exact calculations for relevant experimental
conditions. However, the properties of the $\eta$-condensate are essentially identical in higher
dimensions, and we expect that this switch and ramp scheme will work
similarly in 2D and 3D. We also show that the superlattice scheme has strong
advantages over an alternative schemes involving the adiabatic opening
of a harmonic trap~\cite{RoschVojta}. We then discuss how errors in
state preparation or implementation of the Hubbard Hamiltonian can
be revealed and characterised in experiments via measurements made on the
$\eta$-condensate.

\begin{figure}[t]
\begin{center}
\includegraphics[width=8.5cm]{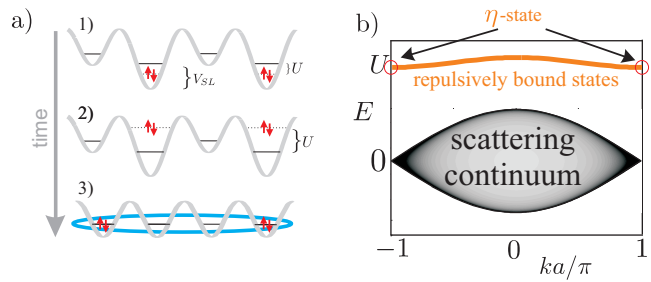}
\caption{a) Preparation of an $\eta$-condensate: 1) Begin in an
insulator state $|\psi_i\rangle$ with attractive onsite interactions
$U$ in an optical superlattice with depth $V_{SL}$; 2) Switch $U$ to
a positive value larger than the bandgap; 3) Delocalise onsite pairs
by adiabatic removal of the superlattice. b) Full spectrum of
energies for $H_{FH}$ with $U>0$ in 1D for a single pair of atoms,
one of each spin species, plotted as a function of centre of mass
quasimomentum $k$, with $a$ the lattice
spacing.}\label{fig:overview2}
\end{center}
\end{figure}

The target state of our switch and ramp process, the
$\eta$-condensate, is an exact excited eigenstate of the Fermi
Hubbard Hamiltonian ($\hbar=1$) in $D$ dimensions
\begin{equation}
\label{HFH} H_{FH}=-J\sum_{\langle {{\bf i}},{{\bf
j}}\rangle,\sigma} \cdag{{{\bf i}},\sigma}c_{{{\bf j}},\sigma}+
U\sum_{{{\bf i}}}n_{{{\bf i}},\upa}n_{{{\bf i}},\dna}.
\end{equation}
This Hamiltonian describes the dynamics of atoms in the lowest band
of an optical lattice~\cite{Reviews, hofstetter}, with $c_{{\bf
i},\sigma}$ a fermionic annihilation operator for particles of spin
$\sigma\,\{\upa,\dna\}$ on lattice site ${\bf i}=(i_1\ldots,i_D)$,
$J$ the tunnelling amplitude, $U$ the onsite interaction energy
shift, and $n_{{\bf i},\sigma}\equiv c_{{\bf i}\sigma}^\dag c_{{\bf
i}\sigma}$. The $\eta$ condensate can be constructed via the
operator $\etd\equiv \sum_{\textbf{i}}
(-1)^{\sum_{d=1}^Di_d}\cdag{{\bf i},\upa}\cdag{{\bf i},\dna}$ first
introduced by Yang, which has the property $[H_{FH},\etd]=U\etd$.
The state $\cond\sim(\eta^\dag)^N \ket{\rm vac}$ is an eigenstate of
$H_{FH}$ with energy $NU$ for positive integer $N$. Below we focus
on the case $U>0$, where $\cond$ is a condensate of $N$ repulsively
bound atom pairs~\cite{RepPairs1}. In Fig.~1b, we plot the
eigenenergies of $H_{FH}$ when we have one particle of each spin on
a 1D lattice, as a function of the centre-of-mass quasimomentum. The
single $\eta$ pair is indicated in the plot, and corresponds to a
repulsively bound onsite pair at the edge of the Brillouin zone,
i.e., with center-of-mass quasimomentum $\pi/a$.

\textit{Switch and ramp process:-} The $\eta$-condensate with $N$ pairs can be prepared using a switch
and ramp process, combining an adiabatic ramp with a sudden switch
in the interaction strength. Adiabatic ramps have previously been
discussed for preparation of many-body ground states in optical
lattices~\cite{Loading1,Loading2}. In an adiabatic ramp, one
prepares a state $\ket{\psi_f}$ of a Hamiltonian $H_0$ beginning
from a non-degenerate, gapped initial state $\ket{\psi_i}$ that is
an eigenstate of the Hamiltonian $H_0 +V$. By removing $V$
adiabatically, the state follows the instantaneous eigenstates of
$H_0+V(t)$ and ends in $\ket{\psi_f}$. The key is that
$\ket{\psi_i}$ should be a gapped state of $H_0 +V$ that is easy to
prepare with very low entropy via standard cooling and loading
techniques \cite{Loading1,Loading3}.

Here we propose to begin from a band insulator in the lowest sites
of an optical superlattice~\cite{Superlattice}, as depicted on the
left in Fig.~2a, which is the ground state of $H_{FH}+V$, with $V$
the Hamiltonian describing the superlattice potential. For the case
depicted in Figs.~1a, 2a, where the superlattice period is twice the
original lattice spacing, $V=V_{SL} \sum_{i\, \textrm{even}} n_i$.
This state has an energy gap $\epsilon_{SL}\sim V_{SL}$
corresponding to the superlattice bandgap, and a filling factor
which is set by the superlattice period~\cite{Loading1} (e.g., half
filling in Figs.~1a, 2a). If we were to let $V_{SL}\rightarrow 0$
adiabatically we would connect this ground state to the ground state
of $H_{FH}$. Instead, we can suddenly switch $U$ (on a timescale
short compared with $J^{-1}$) to a value larger than $\epsilon_{SL}$
(see Fig. 1a). In the limit $|U-\epsilon_{SL}|\gg J$, this switching
will create an excited eigenstate of $H_{FH}+V$, as shown in the
transition from the left panel to the right panel in Fig.~2a.
Adiabatic removal of the superlattice, $V_{SL}\rightarrow 0$, will
then lead to an excited eigenstate of $H_{FH}$. This latter state
will correspond to the lowest energy state in which all particles
exist in repulsively bound pairs, which is the $\eta$-condensate.
The energy spectrum of the Hamiltonian for a small system in 1D is
plotted during the ramp in Fig.~2a, and we see that the state is
always separated by a gap $\sim U$ to lower lying states, and the
gap from the superlattice $\sim V_{SL}$. Whilst in general, the
adiabatic ramp could be optimised using optimal control
methods~\cite{BorziHohenester}, we choose a simple exponential ramp
for the superlattice, $V_{SL}(t)=(e^{-t\nu}-e^{-\nu T})/(1-e^{-\nu
T})$, motivated by the approximate linear dependence of the gap on
$V_{SL}$. Here, $T$ is the total ramp time, and $\nu$ the ramp
speed. Note that for large system sizes, the energy gap can become
very small, but that for finite systems, a gap will always exist to
the other excited states. The key question is how slow this ramp
should be in order to obtain the $\eta$-condensate with high
fidelity for realistic system sizes $\sim 100$
sites~\cite{LatticeSizeReference}.

\begin{figure}[t]
\begin{center}
\includegraphics[width=8.5cm]{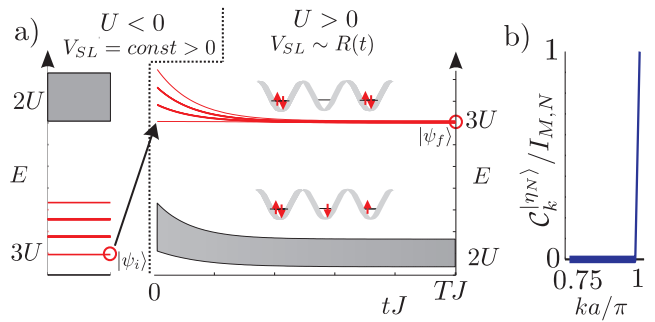}
\end{center}
\caption{a) Energy eigenstates of $H_{FH}+V$ as a function of time
for a small example system with $N_{\upa}=N_{\dna}=3$, $M=6$. Left:
Lowest energy states for strong initial attraction $U/J=-30$, in the
presence of a superlattice. The lowest energy state is the initial
state in our preparation scheme, $|\psi_i\rangle$. The shaded area
denoted the excited manifold of states with one dissociated pair.
Right: The highest energy levels of $H_{\rm FH}+V(t)$ as a function
of time during the adiabatic ramp. The lowest energy eigenstate
$|\psi(t)\rangle$ in the upper manifold where all atoms exist in
pairs is equal to $\cond$ at $t=T$. During the ramp, a gap of order
$U$ always exists to the manifold (shaded area) where some atoms are
unpaired, and a gap to higher levels is present when the
superlattice is present. b) Pair momentum distribution
$\mathcal{C}^{\cond}_k$ of a perfect $\eta$-condensate (see
text).}\label{fig:overview1}
\end{figure}

\textit{Fidelity measures:-} We measure closeness of the final state $|\psi_f\rangle$ to the
$\eta$-condensate in two ways: a) Via the full many-body fidelity
$\mathcal{F}\equiv|\langle\psi_f\cond|^2$, and b) via the similarity
of characteristic correlation functions of $|\psi_f\rangle$ to those
of the $\eta$ state. Remarkably, we will show below that fidelities
$\mathcal{F}\sim 1$ can be obtained for long ramps, despite the fact
that this quantity is exponentially sensitive to the system size, due to the increase
in the size of the many-body Hilbert space.
Indeed, we note that in large systems, states close to $\cond$ can
have essentially the same physical character as the desired state,
and the associated correlation functions may not be significantly
changed by a few small defects in the state, even if $\mathcal{F}$
becomes small. We thus also consider the comparison between
characteristic correlation functions for the final state and
$\cond$, which gives a measure that can be directly measured in
experiments, and is not eponentially sensitive to the size of the
system. In particular, we are interested in the pair momentum
distribution
$\mathcal{C}_k(t)\equiv\mathcal{C}^{|\psi(t)\rangle}_k$, which can
be measured, e.g., by associating atoms in doubly occupied sites to
molecules, and releasing them from the lattice to perform a
time-of-flight measurement. This correlation function is strongly
peaked for $\cond$, reflecting the off-diagonal long-range order
(ODLRO) exhibited by the $\eta$-condensate in any dimension, with
the pairing correlator
$\label{bcscor}C^{\cond}_{\textbf{m},\textbf{n}}=\dnoc
\cdag{\textbf{m},\upa}\cdag{\textbf{m},\dna}
c_{\textbf{n},\dna}c_{\textbf{n},\upa}\cond={I_{M,N}}e^{i\pi(\textbf{m}-\textbf{n})/M}$,
(if $\textbf{m}\neq \textbf{n}$), and $I_{M,N}\equiv N(M-N)/(M-1)$.
The pair momentum distribution is the Fourier transform of this
quantity, $\mathcal{C}_{\textbf{k}}^{\cond}\equiv
\sum_{\textbf{m},\textbf{n}}e
^{i\textbf{k}(\textbf{m}-\textbf{n})}C^{\cond}_{\textbf{m},\textbf{n}}=I_{M,N}\delta_{\textbf{k},\pm\pi/a}$
(see Fig~\ref{fig:overview1}b). We will also consider the total
distribution distance $\mathcal{D}(t)\equiv
1-\sum_{\textbf{k}}|\mathcal{C}_{\textbf{k}}(t)-\mathcal{C}^{\cond}_{\textbf{k}}|/
\sum_{\textbf{k}}|\mathcal{C}_{\textbf{k}}(t)+\mathcal{C}^{\cond}_{\textbf{k}}|$.

\begin{figure}[t]
\begin{center}
\includegraphics[width=8.5cm]{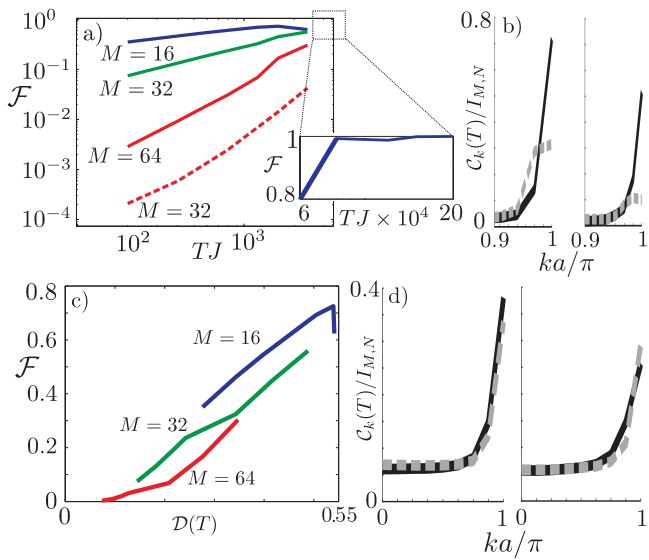}
\caption{a) Fidelities for the superlattice and parabolic trap ramp
as a function of ramp time $T$, computed using $H_{XY}$. The
superlattice ramp shape is $V_{SL}=2J (e^{-\nu t}-e^{-\nu
T})/(1-e^{-\nu T})$, $\nu=J/8$. For the parabolic trap, we use the
same shape with initial $V_{P}/J=0.1$, $\nu=J/12$. $U$ is decreased
with the same shape as the potential in each case, with $U=30J$ at
$t=0$. The inset shows results for longer ramp times with $M=16$.
b) Onsite pair momentum distribution after $T=2400J^{-1}$ for the
superlattice (solid lines) and parabolic trap (dotted lines) ramps
for $M=32$ (left) and $M=64$ (right), computed using $H_{XY}$. c)
Final state fidelity $\mathcal{F}$ as a function of correlation
function distance $\mathcal{D}(T)$ from the perfect
$\eta$-condenstate, computed using $H_{XY}$. d) $\mathcal{C}_k(T)$
for superlattice ramps, computed using $H_{FH}$, with a number of
impurities $N_i=1$ (left) and $N_i=2$ (right),  for $T=200J^{-1}$
(solid black), and $T=400J^{-1}$ (grey dashed).}\label{fig:finalfid}
\end{center}
\end{figure}

\textit{Many-body Fidelities:-} In Fig.~3a we plot the fidelity
$\mathcal{F}$ at the end of the ramp as a function of ramp time $T$
for different system sizes $M$. In order to perform more accurate
calculations for reasonable computational time, these results are
obtained in the limit $U\gg J$. On states that have only repulsively
bound pairs, Hamiltonian~(\ref{HFH}) acts as
$H_{XY}=-(J^2/U)\sum_{\langle
i,j\rangle}\bold{S}_{i}\bold{S}_{j}+2V_{SL}\sum_{i\,
\textrm{even}}S_{i}^z$ in second order perturbation theory, with
$\textbf{S}_i=(S_i^x,S_i^y,S_i^z)$ denoting a vector of spin-$1/2$
operators, and spin states corresponding to sites that are occupied
or unoccupied by a pair of atoms \footnote{We obtain good agreement
when comparing the values we obtain to simulation of the full
Hamiltonian for smaller system sizes.}. Remarkably, for long ramp
times it is possible to obtain unit fidelity, i.e., essentially
perfect $\eta$-condensates. The fidelities are also high for typical
experimental sizes and shorter timescales, with $M=64$, $T\lesssim
1000 J^{-1}$. Although the timescales required to obtain a fixed
fidelity increase with system size, we note (i) that we are already
in the regime of experimentally relevant system sizes, and (ii) that
the sensitivity of $\mathcal{F}$ increases exponentially with the
size of the system, as discussed above.


\textit{Pair momentum distributions:-} This picture is complemented
by the pair momentum distributions, depicted in Fig.~3b. In each
case, the $\eta$-pairing peaks $\mathcal{C}_{\pi/a}$ are clearly
visible, though for ramps with final fidelity lower than one, these
peaks are somewhat broadened. In Fig.~3c we quantify this
relationship between the fidelity $\mathcal{F}$ and the overlap of
the pair momentum distribution with that of the perfect
$\eta$-condensate, as measured by $\mathcal{D}(T)$.  Over a wide
range of $T$ and for different system sizes, we see that these
quantities are strongly correlated, so that sharpness of the peak
could be used to infer the quality of the $\eta$-condensate in
experiments.

\textit{Comparison with opening a harmonic trap:-} For the same
range of $T$ and $M=32,48,64$ we also compare our superlattice
scheme to an adiabatic preparation scheme that was recently
proposed, in which a band insulator is formed in the centre of a
harmonic trap $V_{ \rm trap}\equiv \sum_i V_P(ia)^2$, and the trap
is then opened to produce the final state~\cite{RoschVojta}. As
shown in Fig.~3a, we see that for the same system sizes and ramping
times, we obtain fidelities that are roughly two orders of magnitude
smaller from ramping the harmonic trap. For $M>32$, we see poorer
scaling for the harmonic trap ramps than for the superlattice ramp
(For $64$ lattice sites we obtain fidelities $\mathcal{F}\sim
10^{-12}$). Similar effects are seen in Fig.~3b in the broadening of
the final pair momentum distribution. The superlattice ramp appears
to have an advantage over the harmonic trap scheme because the atoms
do not need to tunnel across the whole system during the ramp, but
rather establish coherence locally.

\textit{Imperfections:-} We now investigate imperfections in the
state preparation process. We will start by addressing how missing
atoms in the initial state, noise, and harmonic trapping potentials
affect preparation of the $\eta$-condensate. We will then discuss
how measurement of time-dependence of correlation functions for the
final state can be used to reveal and characterise these
imperfections in experiments.

\textit{Imperfections - missing atoms:-} To study the impact of
missing atoms in the initial insulator state, we computed the
time-evolution of the adiabatic ramp (with the full Hamiltonian)
starting with localised defects. Regardless of where these defects
are present, and whether we have only missing atoms or complete
missing pairs, this results in a broadening of the peaks in the pair
momentum distribution. Examples are shown in Fig.~3d for a ramp at
half filling with a number of missing atoms $N_i=1,2$. The resulting
correlation functions are, however, stable in time (see below for
further discussion).

\textit{Imperfections - noise:-} Motivated by recent discussions
\footnote{T. L. Ho, private communication}, we also investigated
this ramp in the presence of noise. This would primarily arise from
fluctuations in the lattice depth, which would change the value of
$J$. Note that in the superlattice ramp, $J$ (coupling neighbouring
sites) is always non-zero, even though the effective tunnelling at
the beginning of the ramp is made small by the superlattice, $\sim
J^2/\epsilon_{SL}$. With a variation of $J$ up to 10\% with a
variety of correlation times for the noise, we found no significant
effect on the final state fidelity.

\textit{Effect of a harmonic trap on preparation in a
superlattice:-} If a harmonic trap $V_{\rm trap}$ is present for the
duration of the preparation, we find that the character of the final
state in terms of the pair momentum distribution is close to the
$\eta$-condensate, though the peaks are slightly broadened and the
density profile will correspond to a trap. This state is close to an
excited eigenstate in the presence of the trap, and for $U\gg J$ is
well approximated by an ansatz
$\etd_{\textbf{A}}=\sum_iA_ic^\dag_{i,\uparrow}c^\dag_{i,\downarrow}$,
where $\textbf{A}\in\mathbb{R}^M$ correspond to the ground state
wavefunction of a \emph{single} bound particle with tunnelling
amplitude $-J^2/U$ in the presence of a trapping potential $2V_{\rm
trap}$.

\begin{figure}[t]
\begin{center}
\includegraphics[width=8.5cm]{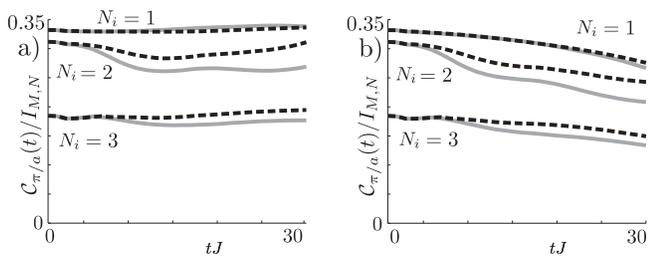}
\caption{Stability of a state close to an $\eta$-condensate with imperfections. a) Time dependence of $\mathcal{C}_{\pi/a}$ for the initial
state with impurities as defined in the text on 32 sites, with 16 onsite pairs and varying
impurity count $N_i$, for $U/J=4$ (dashed black), $U/J=10$ (solid grey). b) Same as (a), but with additional trapping potential $V_P/J=1.25\times10^{-4}$.
}\label{fig:stab}
\end{center}
\end{figure}

\textit{Revealing and characterising imperfections with an
$\eta-$condensate:-} The $\eta$-condensate is an exact excited
eigenstate of the Fermi Hubbard model, and the correlation functions
will be both sharply peaked and stationary, unless there are errors
in the state preparation or implementation of the Hamiltonian.
Broadening and time-dependence of the correlation functions can be
used to reveal imperfections, and also to characterise their source.
We consider an initial eta-state, with $N_i$ delocalised impurity
atoms (see below for more details), and in Fig.~4 we plot the time
dependence of the height of the peak in the pair momentum
distribution. In Fig.~4a we consider only the additional atoms, and
in Fig.~4b we add also a weak additional harmonic trapping
potential. As in Fig.~3d, increasing $N_i$ reduces the height of the
$\eta$-pairing peak. However, provided $U\gtrsim 4J$, the resulting
pairs, and the correlation functions are stable as a function of
time. For $U<4J$ (not shown) the pairs can decay through collision
with unpaired atoms ~\cite{RepPairs2}, and the peak in the pair
momentum distribution also decays. On the other hand, additional
potentials will dephase the state, and cause decay of the peak, as
shown for a very weak harmonic trapping potential in Fig.~4b. The
rate of decay is larger for stronger traps due to faster dephasing,
and unlike the effect of missing atoms, is independent of $U/J$.
This difference could be used in an experiment to characterise the
source of defects in the final state. Note that in order to make
this discussion independent of the form of the ramp, we have
obtained the results in Fig.~4 beginning from
 a state of the form
$\ket{\et,N,N_i}\equiv\sum_{\{i\},\{j\}} :\prod_{n=1}^N(-1)^{i_n}\etd_{i_n}\prod_{k=1}^{N_i}e^{i\delta
j_k}c^{\dagger}_{j_{k},\dna}:\ket{0}$ where $:\ldots:$
denotes the ordering operator by site, i.e
$:\etd_xc^{\dagger}_{y,\dna}:=\etd_xc^{\dagger}_{y,\dna}$ if $x<y$
and $=c^{\dagger}_{y,\dna}\etd_x$ otherwise. Note that
$|\eta,N,N_i=0\rangle=\cond$.

\textit{Outlook:-} The preparation of the $\eta$-condensate offers a
testbed to verify the emulation of many-body Hamiltonians in optical
lattices, providing both a sensitive means to validate the
implementation of the Hamiltonian, and also an important test case
for state preparation schemes involving adiabatic ramps. These
schemes are particularly important in light of the current
experimental challenge to reduce entropies in order to generate
states such as an anti-ferromagnetic phase of the Fermi-Hubbard
model \cite{StoofBloch}.


We thank H. P. B\"uchler for discussions. This work was
supported by the Austrian Science Foundation (FWF) through SFB F40
FOQUS and project I118\_N16 (EuroQUAM\_DQS), the DARPA OLE program
and by the Austrian Ministry of Science BMWF via the
UniInfrastrukturprogramm of the Forschungsplattform Scientific
Computing and Centre for Quantum Physics.

\appendix

\end{document}